\documentclass[runningheads]{llncs}
\usepackage[T1]{fontenc}
\usepackage[utf8]{inputenc}
\usepackage{amsmath,amssymb}
\usepackage{todonotes}
\usepackage{url}
\usepackage{hyperref} 
\usepackage[margin=1in]{geometry}
\usepackage{listings}
\usepackage{multirow}
\usepackage{caption}
\usepackage{subcaption}
\usepackage{cite}
\usepackage{color}
\usepackage[capitalise]{cleveref}

\begin{document}
\title{Optimizing the Production of Test Vehicles using Hybrid Constrained Quantum Annealing}
%
%
\author{Adam Glos\inst{1}\orcidID{0000-0001-6320-7699} \and
Akash Kundu\inst{1,2}\orcidID{0000-0002-3540-1061} \thanks{corresponding author, \email{akundu@iitis.pl}} \and
\"{O}zlem~Salehi\inst{1}\orcidID{0000-0003-2033-2881
}}
%

%
\institute{Institute of Theoretical and Applied Informatics, Polish Academy of Sciences, Bałtycka 5, 44-100 Gliwice, Poland
\and
Joint Doctoral School, Silesian University of Technology, Akademicka 2A, 44-100 Gliwice, Poland}
\maketitle              
\begin{abstract}
		Optimization of pre-production vehicle configurations is one of the challenges in the automotive industry. Given a list of tests requiring cars with certain features, it is desirable to find the minimum number of cars that cover the tests and obey the configuration rules. In this paper, we model the problem in the framework of satisfiability and solve it by utilizing the newly introduced hybrid constrained quadratic model (CQM) solver provided by D-Wave. The problem definition is based on the ``Optimizing the Production of Test Vehicles'' use case given in the BMW Quantum Computing Challenge. We formulate a constrained quadratic model for the problem and use a greedy algorithm to configure the cars. We benchmark the results obtained from the CQM solver with the results from the classical solvers like CBC (Coin-or branch and cut) and Gurobi. We conclude that the performance of the CQM solver is comparable to classical solvers in optimizing the number of test vehicles. As an extension to the problem, we describe how the scheduling of the tests can be incorporated into the model. 

\keywords{Vehicle configuration  \and constrained quadratic model \and quantum annealing \and D-Wave \and BMW challenge.}
\end{abstract}
\section{Introduction}\label{intro}

Quantum computers are deemed promising technologies for solving industrial problems from various sectors like automotive, chemical, insurance, and technology. One of the main problem domains for industrial problems is optimization, as identified in the report prepared by Quantum Technology and Application Consortium (QUTAC) \cite{bayerstadler2021industry}. Recently, there have been attempts to solve optimization problems using near-term quantum computers, through variational quantum eigensolver (VQE) \cite{tilly2021variational}, quantum approximate optimization algorithm (QAOA) \cite{farhi2014qaoa}, and quantum annealing (QA) \cite{das2008annealing}.

Quantum annealing is a heuristic method for solving optimization problems. It operates in the framework of quantum adiabatic computing, which is a quantum computing model alternative to gate-based. Since many optimization problems are proven to be NP-hard, quantum annealing has gained significant interest as an up-and-coming tool to target them. Quantum annealers are commercially available by the D-Wave company \cite{johnson2011quantum}, and a vast amount of research has been devoted to identifying potential use cases \cite{applicationdwave}. D-Wave quantum annealers have been utilized to solve problems from different domains such as transportation \cite{domino2021quadratic, salehi2022unconstrained, domino2021quantum, yarkoni2021multi}, finance \cite{mugel2022dynamic, kurowski2020hybrid}, chemistry \cite{genin2019quantum, teplukhin2020electronic, mato2021quantum}, and computer science \cite{asproni2020accuracy,jiang2018quantum}.

Identified among the use-cases of quantum computing by BMW Group \cite{luckow2021quantum}, optimization of pre-production vehicle configurations is one of the challenges in the automotive industry. Every year, new features and car components are launched by the companies, and various tests should be carried out before the series production. The tests under consideration range from the validation of the model's functionality to the evaluation of the new components. Consequently, pre-production vehicles are built for testing purposes. As the construction of pre-production vehicles is costly and complex \cite{tiepelt2016finding}, it is desirable to reduce the number of required test vehicles. Hence, the test cars should be configured to cover as many tests as possible while meeting some dependency constraints among the different features.   

Some of the attempts in solving the test-vehicle configuration optimization problem use the framework of satisfiability. The features of the vehicle are represented by Boolean variables, indicating whether the feature or the component exists or not. As each vehicle configuration should satisfy the feasibility rules concerning the different features of the car and the requirements imposed by the tests, such rules can be modeled through Boolean constraints. In \cite{tiepelt2016finding}, the authors present a Max-SAT framework that uses a greedy approach and tests it on small-scale real-world data. The problem is also studied through finding minimum set cover in \cite{walter2015optimal}, where the authors formulate the problem as a minimum set cover problem and use SAT solver to check the feasibility of the configurations.  

In this paper, our main goal is to exploit quantum annealing to solve the test-vehicle configuration problem. Our problem definition is based on the use case ``Optimizing the Production of Test Vehicles'' given in the BMW Quantum Computing Challenge \cite{bmwchannelge} and takes into account various buildability constraints. We use an optimization approach, where all the variables are Boolean, and the conditions are given through Boolean constraints, and we aim to minimize the number of vehicles that will be used in testing. Among the various solvers provided by D-Wave, we use the newly introduced hybrid solver for constrained quadratic models \cite{cqm}. The mentioned hybrid solver requires the problem to be encoded as a constrained quadratic model (CQM). In CQM, which is also known as the quadratically constrained quadratic programming in the literature, the problem is identified through a quadratic objective function and quadratic constraints defined over binary and integer variables.  Once the problem is formulated, the hybrid solver takes advantage of both the classical heuristic methods and the D-Wave quantum processors.  As the hybrid solver is the proprietary of the D-Wave company, the exact way it operates is not revealed.

While modeling the problem as a CQM, our primary concern is to use as few qubits as possible. We propose an optimization model for which the total number of qubits grows in the order $O(n(f + o + q ))$, where $n$, $f$,  $o$, $q$ are the numbers of vehicles, features, vehicle types, and tests respectively, and the number of required qubits is independent of the number of constraints. The analysis applies for both SAT and Max-SAT formulations, the former answering the question ``Are $n$ cars sufficient to cover all the tests?", and the latter aiming to `maximize the number of tests covered using $n$ test vehicles`. To benchmark the results obtained from the CQM solver, we develop an integer linear programming formulation. Since both problems are notably time-consuming for current quantum and classical solvers, in order to benchmark the efficiency of the classical and quantum solvers, we analyze the performance of a greedy optimization procedure based on the Max-SAT formulation. We test the performance of the algorithms on the dataset provided by the BMW Quantum Computing Challenge~\cite{bmwchannelge} using D-Wave's hybrid CQM Solver, and the classical solvers CBC (COIN-OR Branch-and-Cut) and Gurobi. The results indicate that the classical optimization algorithms outperform the performance of D-Wave's hybrid solver in minimizing the number of required cars and running time. Furthermore, we consider a variation of SAT problem that includes the scheduling of the tests, however, this model requires far more qubits; thus we claim that it is particularly inefficient for practical purposes. As per the authors' knowledge, this is the first attempt to benchmark the performance of the hybrid solver for CQM compared to classical solvers.

The rest of the paper is organized as follows. In Sec.~\ref{sec:preliminaries}, we present basic concepts related to SAT, Max-SAT, linear programming, quantum annealing, and constrained quadratic model as well as discuss the problem statement. In Sec.~\ref{sec:problem-formulation} we describe the problem formulation and the optimization approach we follow for solving the problem including a brief discussion on the required resources. In Sec.~\ref{sec:results}, we present and discuss the experimental results. We conclude with final comments in Sec.~\ref{sec:conclusion}.

\section{Background} \label{sec:preliminaries}

In this section, we briefly explain the necessary background information on satisfiability problems, linear programming, and quantum annealing concepts.

\subsection{Satisfiability problems} 

\textit{Satisfiability problem} \cite{gu1996algorithms} (SAT) is the problem of determining whether there exists an assignment to the binary variables that make a given Boolean expression true. A \textit{Boolean expression} consists of \textit{Boolean variables}  $x_1,x_2,\ldots, x_n$, that are combined together using logical OR ($\lor$) and AND ($\land$) operations and the negation operator ($\neg$). For instance, the Booelan expression $\phi = (\neg x_1 \land x_2) \lor x_3$ is satisfiable as $x_1=0, x_2=1, x_3=0$ is a satisfying assignment for $\phi$. SAT is the first problem to be proven to be NP-Complete~\cite{cook1971complexity, garey1979guide}. Hence, solving a large family of problems recognized as core to a number of areas in computer science and mathematics is as hard as solving the SAT problem.

A formula is said to be in \textit{conjunctive normal form} (CNF), if it is written as conjunction of clauses. A \textit{clause} is a disjunction of \textit{literals}, where a literal is either a variable (positive literal) or its negation (negative literal). Any Boolean formula can be expressed in CNF. For instance we can express $\phi$ in CNF as $(\neg x_1 \lor x_3) \land (x_2 \lor x_3)$. SAT problem can be equivalently defined as the question of whether there exists an assignment to the Boolean variables that make the formula
\begin{equation}
	C_1\land C_2\land C_3\land\ldots\land C_m = \bigwedge_{i=1}^m C_i
\end{equation}
satisfiable, where $C_i$ is a clause and there are $m$ clauses in total. 

\textit{Maximum satisfiability problem} (Max-SAT) is a generalization of the SAT problem and can be considered as the optimization variant of the decision version. The goal is to find an assignment that maximizes the number of satisfied clauses. Note that knowing the optimal number of satisfied clauses, we can also deduce the solution to the SAT problem; therefore, we can conclude that the Max-SAT problem is NP-Hard. 

A further generalization is the \textit{weighted maximum satisfiability problem} (weighted Max-SAT), in which each clause is associated with a weight, and the aim is to maximize the weighted sum of the satisfied clauses. 

\subsection{Linear programming}

\textit{Linear programming} (LP) is concerned with optimization of an objective function subject to equality and inequality constraints, such that the objective and the corresponding constraints are linear. Linear programs can be expressed in canonical form as
\begin{alignat*}{3}
	&\text{minimize} \hspace{1em}&& c^\intercal x \\
	&\text{subject to} \hspace{1em }&&Ax \leq b \\
	&{} &&x \geq 0,
\end{alignat*}
where $c \in \mathbb{R}^n$, $A \in \mathbb{R}^{n \times n}$ and $ b \in \mathbb{R}^n$. The aim is to find the vector of variables $x$ that minimizes the objective function subject to the given constraints. When all variables $x_i$ are integers, then the problem takes the name \textit{integer linear programming} (ILP) which is NP-Hard in general. If the variables are further restricted to the set $\{0,1\}$, then the problem is called \textit{0-1 linear programming} (0-1 LP). Any ILP can be converted into 0-1 LP. For solving ILPs, there are heuristic methods like simulated annealing \cite{kirkpatrick1983optimization} and exact methods including cutting plane \cite{kelley1960} and branch-and-bound methods \cite{Lawler1966}.

There are various commercial and open-source solvers and toolkits for solving linear programs. PulP is a Python library \cite{mitchell2011pulp} that provides tools for modeling problems and an interface for accessing various solvers. In this paper, we use PulP to model our problems and two different solvers to get the results: CBC (Coin-or branch and cut)\cite{forrest2005cbc} solver which is the default one in PulP, and Gurobi solver \cite{gurobi}.   

\subsection{Quantum annealing}
\textit{Adiabatic quantum computing} (AQC) is an analog computational model that relies on quantum adiabatic theorem which states that a quantum state that is initially in the ground state is likely to stay in the ground state given that the evolution takes place slow enough. Some assumptions of AQC are lifted in \textit{Quantum annealing} (QA), which is a meta-heuristic method for solving optimization problems using the quantum adiabatic theorem.

The problem Hamiltonian $ H_p $ is designed so that its ground state encodes the solution to the problem of interest, and an initial Hamiltonian $ H_0 $ is picked whose ground state is known and easy to prepare. The system is initialized with the ground-state of $ H_0 $, and an adiabatic evolution path is followed so that the system ends up in the ground-state of $ H_p $. This is achieved by evolving the system with the time-varying Hamiltonian $ H(t) $ expressed as  
\begin{equation}
	H(t) = \left(1- \frac{t}{\tau}\right)H_0 + \frac{t}{\tau}H_p.
\end{equation}
The quantum adiabatic theorem assures that the system always remains at the ground state of $H(t)$ for sufficiently large real evolution time $\tau$. When $t=\tau$, $H(t)$ equals $H_p$ and ideally the system is expected to be in the ground state of $ H_p $.	

Commercially available quantum annealers are provided by the D-Wave company. D-Wave quantum processing units (QPUs) implement the initial Hamiltonian $H_0 = -\sum_i \sigma_i^x$, where $\sigma_i^x$ is the Pauli-X operator acting on $i$-th qubit and the problem Hamiltonian should be stated in the form of an Ising model $H_p = \sum_{i>j}J_{ij}\sigma^z_i \sigma^z_j + \sum_{i}h_{i}\sigma^z_i $, where $ J_{ij} $ denotes the interaction between sites $ i $ and $ j $, $ h_{i} $ is the external magnetic field applied on site $ i $ and $\sigma_i^z$ is Pauli-Z operator. Nevertheless, it is much more convenient to formulate problems over binary variables. Any problem that is expressed as a quadratic unconstrained binary optimization (QUBO) problem can be easily converted into an Ising model by replacing the binary variables $b_i$ with $(1-s_i)/2$. QUBO involves the minimization of a quadratic objective function defined over binary variables. While it is an unconstrained model, by using the \textit{penalty method}, one can incorporate the constraints to the objective function \cite{lucas2014ising}. We would like to point out that any ILP formulation can be formulated as a QUBO and we refer readers to \cite{salehi2022unconstrained} for a detailed explanation.  

When running a problem on D-Wave QPUs, the variables need to be mapped to the QPU architecture as the underlying graph representing the interactions in the QPU is not fully connected; this process is known as the minor embedding \cite{choi2008minor}. Hence, the number of variables in the QUBO formulation should be much smaller than the actual number of physical qubits, making it unachievable to run real-world problems when considering the fact that D-Wave Advantage QPU has 5640 qubits. 

D-Wave hybrid solvers can solve much larger problem using a classical-quantum hybrid workflow. With the announcement of the new hybrid solver for constrained quadratic models (CQMs), D-Wave's hybrid solver service (HSS) now consists of three different solvers. Binary quadratic model (BQM) solver accepts problems defined in the form of QUBO. One can define problems over discrete variables in an unconstrained form and use the discrete quadratic model (DQM) solver. In this paper, we will use the CQM solver, which is described in more detail in Sec. \ref{sec:cqm}. All solvers in HSS follow the same workflow. After taking the input, classical heuristic solvers that run parallel on the cloud are called. Those solvers have heuristic and quantum modules that send queries to D-Wave Advantage QPU. The responses taken from the QPU are used to guide the classical heuristic process and improve the quality of the solutions obtained so far. Finally, the heuristic solver returns a solution to the user.  

\subsection{Constrained quadratic model} \label{sec:cqm}

\textit{Constrained quadratic model} (CQM) is the name given by D-Wave to the model involving a quadratic objective function and quadratic constraints that are defined over binary or integer variables. In the literature, this is also known as \textit{quadratically constrained quadratic programming}. We can define a CQM as
\begin{alignat*}{3}
	&\text{minimize} \hspace{1em}&& x^\intercal P_0 x + q_0^\intercal x   \\
	&\text{subject to} \hspace{1em }&&x^\intercal P_i x + q_i^\intercal x \leq r_i \hspace{1em}i=1,\dots,m \\
	&{} &&x \geq 0,
\end{alignat*}
where $P_i \in \mathbb{R}^{n \times n}$,  $q_i \in \mathbb{R}^{n}$ and $r_i \in \mathbb{R}$ for $i=0,1,\dots,m$. The goal is to find the vector $x$ that minimizes the objective function, where $x$ consists of binary and integer variables.

D-Wave has recently introduced the hybrid solver for CQM. Unlike the previous hybrid solvers and quantum annealers of D-Wave, the CQM solver natively supports equality and inequality constraints. This is advantageous for problems involving constraints compared to the QUBO, which is the standard formulation that has been used for quantum annealing so far. First of all, there is no need for removing the constraints through the penalty method, which in turn increases the number of variables if the constraints are in the form of inequalities. Secondly, it removes the difficulty of setting penalty coefficients, which is challenging as the model becomes sophisticated. Thirdly, CQM allows the inclusion of quadratic constraints directly into the model, which is not possible for QUBO. It is also suggested by D-Wave that the hybrid CQM solver should be preferred over the other hybrid solvers in case the problem naturally involves constraints. An experimental evidence for performance comparison is available in \cite{cqm}.

\subsection{Problem definition}

In this section, we will describe the details of the vehicle optimization problem. The configuration of each vehicle is determined by the presence or absence of the availability of the features, and each vehicle has a specific type. Let us discuss the buildability constraints that a vehicle configuration should satisfy \cite{bmwchannelge}.

\begin{itemize}
	\item \textbf{Single type requirement:} Each vehicle should have a single type.
	\item \textbf{Features allowed per type:} For a given type, only some of the features are available.
	\item \textbf{Group features:} Certain groups of features cannot be implemented together (i.e. at most one can be implemented). This constraint is valid for all types.
	\item \textbf{Rules per type:} For each type, there are rules which govern the feature set that the vehicle of a specific type should have. Those are mainly implication rules. 
\end{itemize}

Finally, we have the \textbf{test requirements} that define the properties of the cars needed for the testing phase. We will assume that each test requires a single car and discuss how this assumption can be removed later on. 

Consider $n$ test vehicles, a list of $ f $ features and assume that there are $o$ different types available. Let us assume that there are $ q $ test requirements. For vehicle $i \in [n]$, we will represent the presence/absence of feature $j \in [f]$ through the binary variables $b_{i,j}$, where $b_{i,j}=1$ iff vehicle $i$ has feature $j$. Next we represent the type of the vehicle using the variable $t_{i,j}$, where $t_{i,j} = 1$ iff vehicle $i \in [n]$ is of type $j \in [o]$. And finally we define the binary variables $ p_{i,j}$, to represent whether vehicle is used in test, where $p_{i,j} = 1$ iff vehicle $i\in [n]$ is used for test $j$.

Suppose that there are $ c $ buildability constraints in total. Note that the same set of constraints applies to each vehicle. Each constraint $\phi_k$ can be expressed using a Boolean expression over the binary variables we have defined above. First of all, we would like all buildability constraints to be satisfied for any given number of cars. This can be expressed using the following logical expression:
\begin{equation} \label{eq:build}
	\bigwedge_{i=1}^n \bigwedge_{k=1}^c \phi_k(b_{i,1},\dots, b_{i,f},t_{i,1},\dots, t_{i,o}). 
\end{equation} 
Similarly, we can express the test requirements using Boolean expressions. To start with, each test requires absence or presence of certain features. We need constraint $ \psi_l $ to ensure that the variable $ p_{i,l} $ representing the $ l $'th test requirement is set correctly. 
\begin{equation} \label{eq:testc}
	\bigwedge_{i=1}^n \bigwedge_{l=1}^q  \psi_l(b_{i,1},\dots, b_{i,f},p_{i,l}). 
\end{equation}
Now we can identify two different problems based on how we interpret test requirements. Given $ n $ vehicles, the first problem is to decide whether there exist configurations for the given cars so that the buildability constraints are satisfied, and for each test requirement, there is at least one car satisfying the requirement. This results in the following \textit{satisfiability problem}:
\begin{equation} \label{eq:sat}
	\bigwedge_{i=1}^n \bigwedge_{k=1}^c \phi_k(b_{i,1},\dots, b_{i,f},t_{i,1},\dots, t_{i,o}) \land 	\bigwedge_{i=1}^n  \bigwedge_{l=1}^q \psi_l(b_{i,1},\dots, b_{i,f},p_{i,l}) \land
	\bigwedge_{l=1}^q \bigvee_{i=1}^n  p_{i,l}. 
\end{equation}
If one wants to find the smallest number of vehicles for which the Boolean formula given in Eq.~\eqref{eq:sat} is true, then the bisection method can be used by starting with a large $ n $ and applying binary search to find the optimal value.

It can be the case that the number of vehicles is fixed, and the aim is to find a configuration of vehicles that satisfy the buildability constraints and maximize the number of satisfied test requirements. So, unlike in the case of SAT, it is not required that all the test requirements are satisfied. Furthermore, each test requirement can be assigned some weight, in which case the aim is to maximize the weighted sum of the fulfilled tests. This yields a variant of \textit{weighted maximum satisfiability problem} that is expressed mathematically as follows:
\begin{equation} \label{eq:maxsat}
	\text{maximize}~	\sum_{l=1}^q  w_l \prod_{i=1}^n p_{i,l}, 
\end{equation} 
where $ w_l $ is the weight associated with test requirement $ l $, subject to the constraints
\begin{equation} 
	\bigwedge_{i=1}^n \bigwedge_{k=1}^c \phi_k(b_{i,1},\dots, b_{i,f},t_{i,1},\dots, t_{i,o}) \land 	\bigwedge_{i=1}^n  \bigwedge_{l=1}^q \psi_l(b_{i,1},\dots, b_{i,f},p_{i,l}).
\end{equation} 

\section{ILP formulation}\label{sec:problem-formulation}

Having discussed the general overview for the vehicle testing problem, now we are ready to express it as a linear program defined over binary variables. We will be expressing Boolean expressions that correspond to constraints using equalities and inequalities.

\subsection{Buildability constraints} 	

Here we discuss the formulation of the buildability constraints through linear equalities and inequalities.

\subsubsection{Single type requirement}
The fact that each vehicle should have a single type, can be incorporated using the constraints
\begin{equation}
	\sum_{j=1}^o t_{i,j}= 1, \qquad i\in[n]. \label{eq:constr1}
\end{equation}

\subsubsection{Features allowed per type}
Some of the features are not allowed for a specific type. We encode this constraint through the features which are not allowed for the given type. In other words, for each type $j \in [o]$, there exists a collection of features $F_{j}$ such that $t_{i,j}
=1\implies b_{i,k}=0$ for all $k\in F_{j}$ for all vehicles $i\in [n]$. This is expressed by the inequality constraints
\begin{equation}
	t_{i,j} + b_{i,k} \leq 1, \qquad i\in [n], j\in[o],k \in F_{j}. \label{eq:constr2}
\end{equation}

\subsubsection{Group features}
Let $ F_G $ denote a collection of feature groups. Given a group of features $ G \in F_G $, we need to make sure that at most one feature from each group $ G $ is implemented. This is equivalent to inequalities
\begin{equation}
	\sum_{k\in G} b_{i,k} \leq 1, \qquad i\in [n], \label{eq:constr3}
\end{equation}
for each $G \in F_G$. 

\subsubsection{Rules per type}

Depending on the type of the vehicle, one may define rules in the form of implications about the presence or absence of the features in the vehicle as
\begin{center}
	\texttt{T1: F2 $\land$ $ \neg $F4 $\land$ $ \neg$F5 $\implies$ F1 $\lor$ F3},
\end{center}
where the first term is the type of the vehicle, and on the left and right sides of the implication, we have conjunction or disjunction of literals. The rule is saying that if a vehicle is of type 1, has feature 2, and does not have features 4 and 5, then it should have at least one of the features 1 or 3.

We group the constraints into several classes depending on the form of the constraint and label it with a 4-character string, where the first two characters are representing the LHS of implication, and the remaining two are the RHS. The first character describes whether the variables on the LHS of the implication are negated or not. There are three possibilities: ``\texttt{0}" if none of the variables are negated, ``\texttt{1}" if all variables are negated,
``\texttt{m}" if some of the variables are negated. Second character on LHS describes the operator used: ``\texttt{\&}" for AND, ``\texttt{$|$}" for  OR, \texttt{1} if only a single variable exists. The next two characters have the same meaning but describe the RHS of the constraint. For example, the implication above belongs to the class \texttt{m\&0$|$}. The cases we will consider are inspired by the BMW Quantum Computing Challenge dataset and include the combinations of \texttt{m\&} and \texttt{0|} on the LHS and \texttt{0|}, \texttt{0\&}, \texttt{1|}, \texttt{1\&} on the RHS. We give the logical expression and the corresponding arithmetic expression for the mentioned cases in Table~\ref{tbl:cases}.

\begin{table}[h]
	\centering
	
	\begin{tabular}{ccll}
		\hline
		\multicolumn{1}{c}{Type} & \multicolumn{1}{c}{Position} & \multicolumn{1}{l}{Logical expression} & \multicolumn{1}{l}{\hspace{5pt}Corresponding constraint} \\[7pt] 
		\hline
		\texttt{m\&}  &LHS & $ 	t_{i,j} \land  \bigwedge_{r=1}^M b_{i,j_r} \land \bigwedge_{r=1}^\mu b_{i,l_r} $ & \hspace{5pt} $	1-t_{i,j} + M - \sum_{r=1}^M b_{i,j_r} + \sum_{r=1}^\mu b_{i,l_r}=0$ \\[7pt]
		\texttt{0\&} & RHS & $  \bigwedge_{r=1}^N b_{i,k_r}$ & \hspace{5pt} $N - \sum_{r=1}^N b_{i,k_r}  = 0$ \\[7pt]
		\texttt{1\&} & RHS & $\bigwedge_{r=1}^N \neg  b_{i,k_r}$  & \hspace{5pt} $ \sum_{r=1}^N b_{i,k_r}  = 0 $ \\[7pt]
		\texttt{0|} & RHS	&  $  \bigvee_{r=1}^N b_{i,k_r}$  & \hspace{5pt}  $  1 \leq \sum_{r=1}^N b_{i,k_r}$                \\[7pt]         
		\texttt{1\&} & RHS	&  $  \bigvee_{r=1}^N \neg b_{i,k_r}$  & \hspace{5pt}  $ \sum_{r=1}^N b_{i,k_r}  \leq N-1  $               \\[7pt]   \hline           
	\end{tabular}%
	\caption{Some logical expressions and their corresponding constraints.}
	\label{tbl:cases}
\end{table}

Now we will investigate the implications of each type, making use of the arithmetic expressions given in Table~\ref{tbl:cases}.

\paragraph{Case \texttt{m\&0\&}} This category encapsulates \texttt{m\&01}, \texttt{0\&0\&}, \texttt{1\&0\&}, \texttt{0\&01}, \texttt{1\&01}, \texttt{010\&}, \texttt{0101}, \texttt{110\&}, \texttt{1101}. The constraints take the form
\begin{equation}\label{eq:case1}
	N - \sum_{r=1}^N b_{i,k_r}   \leq N \big(1-t_{i,j} + M -\sum_{r=1}^M b_{i,j_r}+\sum_{r=1}^\mu b_{i,l_r}\big), \qquad i \in [n]. 
\end{equation}
RHS is 0 iff the assumption is satisfied, in which case $ \sum_{r=1}^N b_{i,k_r} $ should be equal to $ N $. If the assumption is not satisfied, then RHS is at least $ N $, hence the inequality is still correct.

\paragraph{Case \texttt{m\&1\&}} This category encapsulates \texttt{m\&11}, \texttt{0\&1\&}, \texttt{1\&1\&}, \texttt{0\&11}, \texttt{011\&}, \texttt{0111} and the reasoning is the same as above
\begin{equation}\label{eq:case2}
	\sum_{r=1}^N b_{i,k_r}  \leq N(1-t_{i,j} + M-  \sum_{r=1}^M b_{i,j_r} + \sum_{r=1}^\mu b_{i,l_r}), \qquad i \in [n]. 
\end{equation}

\paragraph{Case \texttt{m\&0|}} This category encapsulates \texttt{0\&0|}, \texttt{1\&0|}, \texttt{010|}, \texttt{110|}. The constraints take the form 
\begin{equation}\label{eq:case3}
	1 - \sum_{r=1}^N b_{i,k_r}\leq (1-t_{i,j} + M- \sum_{r=1}^M b_{i,j_r} + \sum_{r=1}^\mu b_{i,l_r}), \qquad i \in [n].  
\end{equation}
Note that RHS is 0 iff the assumption is satisfied In this case $ \sum_{r=1}^N b_{i,k_r} \geq 1 $ should be true. If the assumption is not satisfied, then the inequality is still correct. 

\paragraph{Case \texttt{m\&1|}} This category encapsulates \texttt{0\&1|}, \texttt{1\&1|}, \texttt{011|}, \texttt{111|}. The constraints take the form
\begin{equation}\label{eq:case4}
	\sum_{r=1}^N b_{i,k_r} - N + 1 \leq (1-t_{i,j} + M- \sum_{r=1}^M b_{i,j_r} + \sum_{r=1}^\mu b_{i,l_r}), \qquad i \in [n].  
\end{equation}

\paragraph{Case \texttt{0|0|}, \texttt{0|1|}} We take the contrapositives and obtain \texttt{1\&1\&} and \texttt{0\&1\&} respectively, which are already discussed above.

\paragraph{Case \texttt{0|1\&}, \texttt{0|0\&}}
The constraints of the form \texttt{0|1\&} are expressed as  $ \displaystyle
t_{i,j} \land \bigvee_{r=1}^M b_{i,j_r} \implies  \bigwedge_{r=1}^N \neg  b_{i,k_r}
$ and equivalently, we have conditions
\begin{align}\label{eq:case5}
	\centering
	( t_{i,j} \land  b_{i,j_1})  &\implies \bigwedge_{r=1}^N \neg  b_{i,k_r},\\
	&\;\;\;\;\vdots\nonumber\\
	( t_{i,j} \land  b_{i,j_M})  &\implies \bigwedge_{r=1}^N \neg  b_{i,k_r}.
\end{align}
We have $M$ rules of the form \texttt{0\&1\&}. Similarly, the constraints of the form \texttt{0|0\&} translates to rules of the form \texttt{0\&0\&}.

\subsection{Test requirements}\label{sec:requirements}

Test requirements define the properties of cars needed for the testing phase.  Each test requires the absence or presence of certain features. We will assume that the test $ j $ is in the form
\begin{equation}
	b_{i,{k_1}} \land \cdots \land b_{i,k_{T_j^1}} \land \neg b_{i,l_1} \land \cdots \land \neg b_{i,l_{T_{j}^2}} \land ( b_{i,m_1} \lor \cdots \lor b_{i,m_{T_{j}^3}}).
\end{equation}
The values $ T_j^1 ,T_j^2$ and $ T_j^3 $ may be equal to 0. 

Recall that $p_{i,j}$ indicates whether vehicle $i$ is used in test $j$. We need constraints to ensure that the binary variables $ p_{i,j} $ are properly set. Note that test $j$ either imposes some features to exist in the vehicle, in which case we can express it using the inequality
\begin{equation}
	-b_{i,k_r} + p_{i,j} \leq 0, \qquad i \in [n],~ r \in [T_{j}^1], \label{eq:tc1}
\end{equation} 
or imposes that some features should not exist in the vehicle, which results in the inequality
\begin{equation}
	b_{i,l_r} + p_{i,j} \leq 1, \qquad i \in [n],~ r \in [T_{j}^2], \label{eq:tc2}
\end{equation} 
or imposes disjunction of some features which translates as the inequality
\begin{equation}
	p_{i,j}  -\sum_{r=1}^{T_{j}^3} b_{i,m_r} \leq 0 , \qquad i \in [n]. \label{eq:tc3}
\end{equation}
We will call the constraints defined in \cref{eq:tc1,eq:tc2,eq:tc3} as the test constraints. The test constraints are needed both in SAT and Max-SAT approaches.

We will lift the assumption that each test requires only a single car. Let's assume that test $ j $ requires $ k_j $ cars that satisfy the required properties. In case we want to solve the SAT problem, we include the following constraint in our formulation to ensure that the number of vehicles that satisfy test $j$ is $k_j$ for each $j=1,\dots,q$:
\begin{equation}
	\sum_{i=1}^n p_{i,j} = k_j, \qquad j \in [q]. \label{eq:constr11}
\end{equation}

If we want to solve the Max-SAT problem, then we need to define an objective function to maximize. One possibility is to take into account the number of cars that satisfy the test and reflect this in the weight. This results in the following objective function: 
\begin{equation}
	\sum_{i=1}^n \sum_{j=1}^q w_j p_{i,j}.
	\label{eq:partial-objective}
\end{equation} 

It is clear that there should be some upper bound on the number of cars that satisfy a specific test. For example, for a given test $j$, if more than $k_j$ cars satisfy the test, the excess ones should not add to the objective. Hence, we need the following constraint in the case of the Max-SAT approach:
\begin{equation}
	\sum_{i=1}^n p_{i,j} \leq k_j,  \qquad j \in [q].
	\label{eq:kj-bound}
\end{equation}

To conclude, to solve both the SAT problem and Max-SAT problem, we need the buildability and the test constraints defined in \cref{eq:constr1,eq:constr2,eq:constr3,eq:case1,eq:case2,eq:case3,eq:case4,eq:case5,eq:tc1,eq:tc2,eq:tc3}. For the SAT problem, we need additionally the constraint defined in Eq.~\eqref{eq:constr11}. For the Max-SAT problem, we need additionally the constraint defined in Eq.~\eqref{eq:kj-bound} and the objective function is defined as in Eq.~\eqref{eq:partial-objective}.

\subsection{Scheduling}

A related problem in the automotive industry is vehicle test scheduling. Besides the configuration of the vehicles, there are additional constraints about when and in which order the tests will be performed and whether a vehicle can be used in more than one test, making the problem more complicated. The problem has been considered using classical approaches like constraint programming and mixed-integer linear programming in \cite{mitchell2011pulp, forrest2005cbc, shi2017analytical}. From our perspective, the presented model can be extended to incorporate scheduling constraints, as we will discuss briefly.

We assume that each test takes a single day, and the tests should be completed in $D$ days. We extend the previously introduced $p_{i,j}$ variables into $p_{i,j,d}$ where $d\in[D]$ is the day at which test $j$ is performed with vehicle $i$. We replace $p_{i,j}$ in each previously introduced condition with  $p_{i,j,d}$, and if needed a summation over $d$ should be added. For example constraint given in Eq.~\eqref{eq:constr11} will be replaced with 
\begin{equation}
	\sum_{d=1}^D\sum_{i=1}^n p_{i,j,d} = k_j, \label{eq:constr11-scheduling}
\end{equation}
for each test $j\in[q]$.

From now on, we assume that only a single car is needed for each test. If the $j$-th test requires $k_j$ cars, we create variables $p_{ij_1},p_{ij_2},\dots,p_{ij_{k_j}}$ for test $ j $. Hence, the constraint presented in Eq.~\eqref{eq:constr11-scheduling} takes the form
\begin{equation}
	\sum_{d=1}^D\sum_{i=1}^n p_{i,j,d} = 1,
\end{equation}
where $ j $ belongs to the extended list of tests. 

To ensure that test $ j $ is performed within the time frame $[ t_j^{\text start} , t_j^{\text end}]$, we need the constraint
\begin{equation}
	t_j^{\text start} \leq d \cdot \sum_{d=1}^D \sum_{i=1}^n p_{i,j,d} \leq  t_j^{\text end}.
\end{equation}

If for a given test set $\mathcal J=\{j_1,\dots,j_m\}$ we need to use different vehicles for testing, then the vehicle should be used at most once for one of the tests in the given test group. This constraint can be imposed by
\begin{equation}
	\sum_{d=1}^D \sum_{j\in \mathcal J} p_{i,j,d} = 1
\end{equation}
for all $i\in[n]$ and each test set $\mathcal J$.

Let us now consider other conditions on scheduling.	
Let $K$ be the number of cars that can be tested per one day. We need to ensure that for each day $d\in [D]$, the number of tests performed is at most $K$, which is equivalent to 
\begin{equation}
	\sum_{i=1}^n \sum_{j=1}^q p_{i,j,d} \leq K. \label{eq:scheduling-constr1}
\end{equation}
Similarly one has to ensure that each car $i\in[n]$ is
tested at most once in each day $d\in [D]$, which is equivalent to
\begin{equation}
	\sum_{j=1}^q  p_{i,j,d} \leq 1. \label{eq:scheduling-constr2}
\end{equation}

Let us now consider how one can assign groups to each test to impose an order condition among tests from different groups. Let $g_j$ be the group id of the test $j$. We assume, that $g_j$ is an integer in $\{1,\dots,\bar g\}$, s.t. for two tests $j,j'$, where $j$ has to be performed before $j'$ if $g_{j} > g_{j'}$ (\emph{order condition}). In addition, we assume that the tests from group $g_j=1$ are full crash tests, thus not only that they have to be the final test, but also each car can be used only once for such test (\emph{crash condition}).

The order condition can be implemented as follows: For each vehicle $i\in[n]$, for each pair of tests $j,j'$ such that $g_j > g_{j'}$, and for each $d,d'\in[D]$ such that $d<d'$, we add the constraint
\begin{equation}
	p_{i,j',d} + p_{i,j,d'} \leq 1.
\end{equation}
Note that with the approach above, we can handle even more complicated test ordering, like the one defined by partial order of tests \cite{partialorder}.

To implement the crash condition, together with the order condition, it is enough to ensure that the car is used for only one crash test. Let $J_{1}=\{j\in[q]: g_j=1\}$ be the set of tests resulting in a crash. The constraint takes the form for each vehicle $ i \in [n] $
\begin{equation}
	\sum_{j\in J_1} \sum_{d=1}^D p_{i,j,d} \le 1
\end{equation}

\subsection{Resource analysis}
Let us analyze the number of variables and constraints required by the formulation. To start with, there exist $ n\cdot f $ binary variables $ b_{i,j} $, $ n\cdot o $ binary variables $ t_{i,j} $, and $ n\cdot q $ binary variables $ p_{i,j} $. Overall, we need $O( n(f+o+q)) $ binary variables, which grows linearly in the number of vehicles.

There are $ c $ buildability constraints. The number of test requirements depends on the individual tests and can be expressed as $q + \sum_{j=1}^q T_j^1 + T_j^2 + [T_j^3>0]$, where $ [T_j^3>0]=1 $ if $T_j^3>0 $. The first term results either from the constraint Eq.~\eqref{eq:constr11} or Eq.~\eqref{eq:kj-bound}, depending on the problem in consideration. Assuming that  $ T_j^1$ and $ T_j^2$ are negligible compared to $ q $, the total number of constraints can be expressed as $ O(c+q) $. 

Let us now consider the number of qubits used for scheduling constraints. Previous considerations are still valid up to part where $p_{i,j}$s were computed, as they are now replaced with $p_{i,j,d}$. So in total we need $O(n(o+f+qD))$ variables. Note that the polynomial is no longer quadratic.

\section{Experimental results}\label{sec:results}

In this section, we will describe the algorithm, the implementation details and present our results. 

\subsection{Algorithm}
Based on the formulations presented, one can follow different approaches to find the minimum number of required vehicles. The first is the global bisection method, which is also proposed in BMW use-case specification. The idea is to start with a large $ n $ value and then use binary search to find out the optimal $ n $. As the number of variables grows as the product of the number of vehicles and number of constraints, the limitation with this approach is the large number of variable requirements. For instance, in the case of the CQM solver, the number of variables is limited to 5000. In the case of Gurobi, there is no limit on the number of variables and constraints that can be used in principle, however, the time needed for solving the problem increases as the number of variables and constraints increase, which may result in an intractable problem in practice. 

Another approach would be the vehicle-greedy algorithm. We choose an extra parameter $n_{\rm bunch}$ which denotes the number of cars that will be configured at each iteration. After each iteration, the tests are updated by removing the satisfied ones and by diminishing the number of required cars for a test if it is partially satisfied. Then the optimization process is repeated again with the new $n_{\rm bunch}$ cars. The procedure stops after all the tests are covered. Note that we actually maximize the number of covered tests using this approach, and thus solving the Max-SAT problem instance.

\subsection{Implementation}

We used the dataset provided by the BMW Quantum Computing Challenge, which is created based on the BMW Series 2 Gran Coupe. The features and constraints are based on the actual numbers resulting in a real-world problem. The specifications of the dataset are given in Table \ref{tbl:dataset}.

\begin{table}[h]
	\centering
	\begin{tabular}{l@{\quad\quad}ll}
		\hline
		Constraint          & \multicolumn{2}{l}{Number} \\ 
		\hline
		Features allowed per type & \multicolumn{2}{l}{25}                    \\ 
		Rules per type            & \multicolumn{2}{l}{4032}                  \\ 
		Group features            & \multicolumn{2}{l}{41}                    \\ 
		Test requirements          & \multicolumn{2}{l}{643}                   \\ 
		\hline
	\end{tabular}
	\caption{A summary of the number of constraints based on the real-world problem.}
	\label{tbl:dataset}
\end{table}

When analyzing the buildability constraints, we noticed a significant redundancy in the ``rules per type'' constraints. We realized that some of the constraints apply to all types. Secondly, 
some of the constraints apply to all types. In this case, we used a type-independent constraint and heavily reduced the number of constraints. For instance, assuming that the inequality  $b_{i,1} \leq 2-t_{i,j} - b_{i, 2}$ exists for all $ i \in [n] $ and $ j \in [o] $, it can be replaced by $b_{i,1} \leq 1 - b_{i, 2}$ for all $ i \in [n] $. In case some constraint was missing for a number of types, but its inclusion for the remaining types was not disruptive (since the left hand side of the implication could not be satisfied for the particular type with any combination of features), we assumed that it applies for all types and used the discussed simplification. When some features were not available for the given type yet appeared as a positive literal in the constraint, they were removed. Finally, some ``rule per type" constraints possessed redundant information because variables were repeated both on the left and right sides. Those constraints were simplified as well, resulting in new constraint types, which are implemented in a form similar to the ones previously mentioned. 	Besides the buildability constraints, we also performed simplification for the test requirements, by merging the test lines occurring multiple times in the file. Further details can be found in our implementation processing which can be found in the \cite{akash_kundu_2022_6012261} in [path to solver].

We used the vehicle greedy algorithm, taking $ n_{\rm bunch}=1$, hence optimizing a single vehicle at a time. After the mentioned simplifications, the model has 911 binary variables and 6313 constraints. We followed the Max-SAT approach taking Eq.~\eqref{eq:partial-objective} as the objective function and setting all weights equal to 1. We used \texttt{PuLP toolkit} and \texttt{dwave-ocean-sdk} to implement the code for generating the linear program and the constrained quadratic model. The experiments were run on a computer with the following specifications: Intel(R) Core(TM) i9-10900KF CPU @ 3.70GHz; Ubuntu 20.04.3 LTS, 64 GB RAM.

\subsection{Results and discussion}

As mentioned earlier, we obtained the results by setting $n_{\rm bunch}=1$ i.e. one car is configured at each iteration. Hence the number of successful iterations (which outputs a valid car) corresponds to the number of needed vehicles. The algorithm stops if no tests are remaining to be satisfied. For the classical solvers, the experiments are repeated 60 times, and for CQM solver, it is repeated 3 times. CBC and Gurobi algorithms always return the same result in each experiment, i.e. they terminate at the same iteration and return the same number of cars. Meanwhile, for CQM, we took the best possible outcome.

\begin{figure}[tbh!]
	\centering
	\begin{subfigure}{.5\textwidth}
		\centering
		\includegraphics[width=\linewidth]{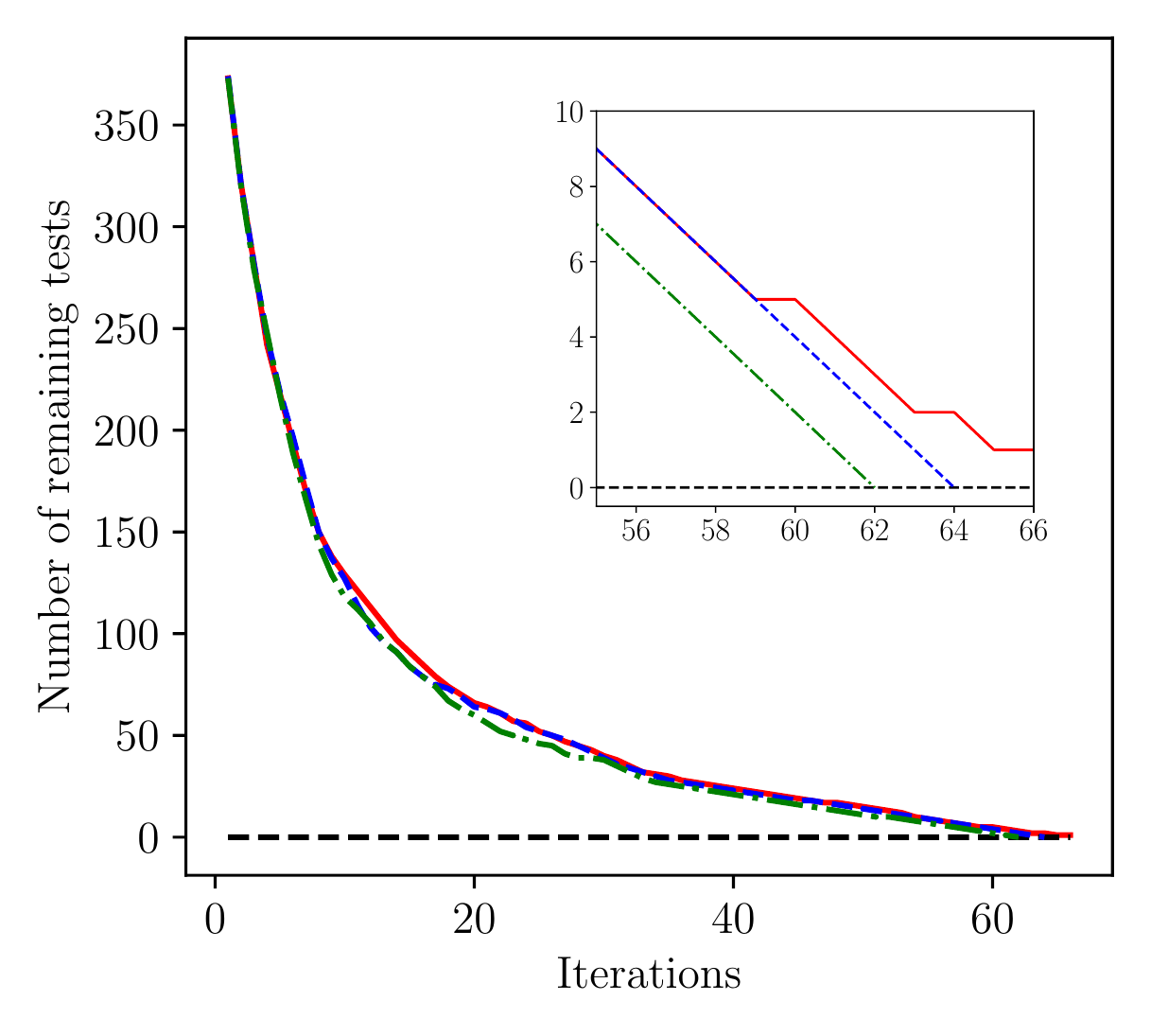}
		\caption{}
		\label{fig:remaining-tests-1-car}
	\end{subfigure}%
	\begin{subfigure}{.5\textwidth}
		\centering
		\includegraphics[width=0.975\linewidth]{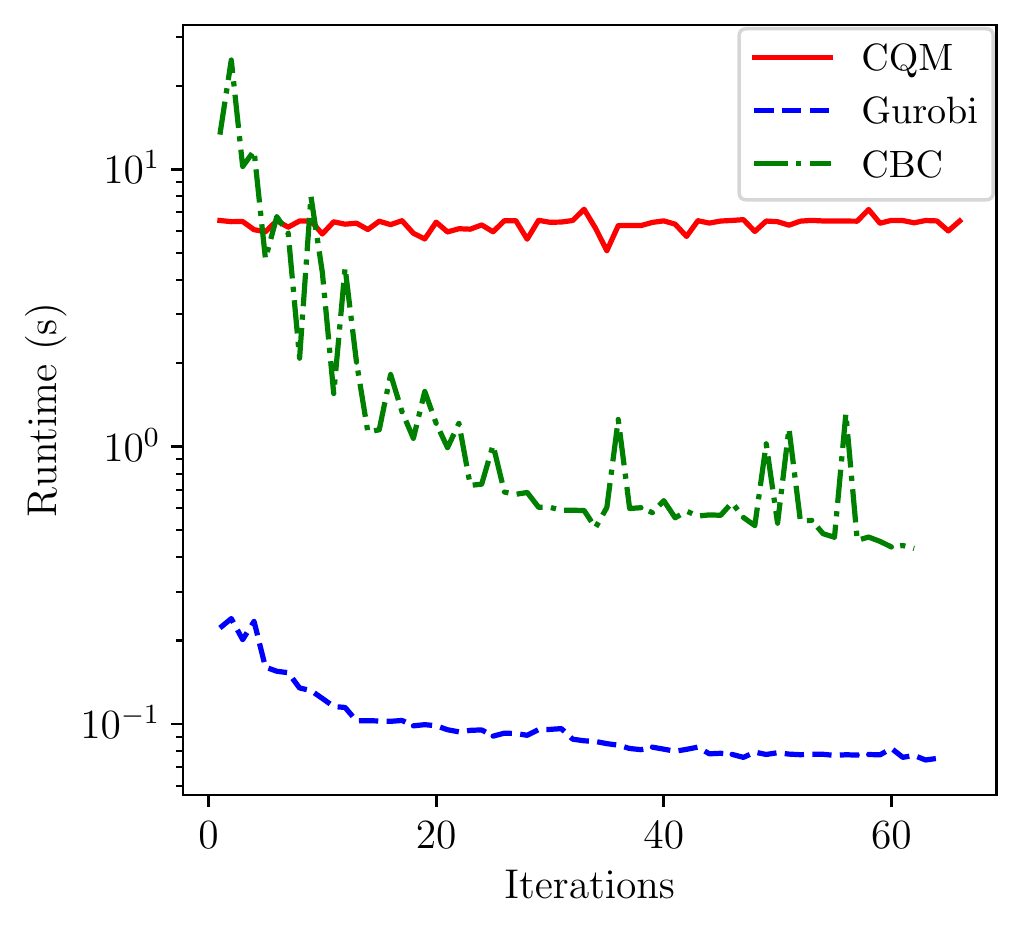}
		\caption{}
		\label{fig:optimizing-1-cars}
	\end{subfigure}
	\caption{Illustration of (a) variation in runtime and (b) number remaining tests with respect to iterations for CQM, CBC and Gurobi solvers.}
	\label{fig:plots-exp-results}
\end{figure}

In the Fig.~\ref{fig:remaining-tests-1-car}, we illustrate the number of remaining tests after each iteration using CQM, CBC, and Gurobi solvers setting $n_{\rm bunch}=1$. CBC solver returns the smallest number of vehicles which is 62, and Gurobi solver returns 64. We would like to note that in the experiment with CQM solver, one test that requires a single vehicle remains after the 65'th iteration. Thus, we can conclude that the CQM solver returns 66; however, the solver fails to find a configuration that satisfies the test in the 66'th iteration.

The problem size gets smaller over the iterations as some tests are removed. We analyze how the runtime changes as the problem size gets smaller in Fig.~\ref{fig:optimizing-1-cars} for $n_{\rm bunch}=1$. The runtimes are calculated by taking the average over the repeated experiments.  The primary observation while comparing the performance of CQM, CBC, and Gurobi is that after the first few iterations, the runtime of CQM solver saturates near $5$ seconds. This is because the default runtime for the solver is $5$ seconds, and one can not go below it.  Meanwhile, the fluctuation in runtime for CBC solver with the number of iterations is visible and varies in the range 1--25 seconds, and the fluctuations in runtime comparatively stabilize for the number of iterations $\ge 30$. Finally, for Gurobi the runtime always stays $\le 1$ second; hence it takes the least amount of time to satisfy all the tests. In Table \ref{tab:averaged-overall-runtime}, a summary of the overall runtime taken by the solvers is depicted. Overall, it takes $6.309$ seconds for Gurobi to find the solution, which is significantly small compared to the other solvers. Although CBC solver takes a longer time, it returns the best solution. CQM solver performs worse both in runtime and minimizing the number of required cars, it takes $371.975$ seconds, in which only $0.001$ of the time is spent on QPU.

We have also checked the performance of the solvers when $ n_{\rm bunch}=5$, which is the maximum possible number that can be taken without exceeding the 5000 variables limit of the CQM solver. Gurobi solver returned 63 cars, so we can say that the result is slightly improved. However, the overall experiment took a significantly longer time (83 seconds). For the CQM solver, no feasible solutions were obtained with the default time limit of 5 seconds. We observed that the returned samples either violated the ``single type constraint" and the vehicles had no type  (in that case all constraints related to ``rules per type" are automatically satisfied) or several other constraints were violated. When the time limit was increased to $10$ seconds, then the CQM solver was able to return a feasible solution at each iteration. However, after the 13'th iteration (after $65$ vehicles were configured), there were still $35$ tests remaining to be satisfied, hence the optimization quality was worse. We would like to remark that the CBC solver failed to return any result within a reasonable amount of time.

\begin{table}[t!]
	\centering
	\begin{tabular}{|c|c|c|c|}
		\hline
		\multicolumn{2}{l}{Solver} & \multicolumn{2}{l}{Overall runtime (in seconds)} \\
		\hline
		\multicolumn{2}{l}{Gurobi} & \multicolumn{2}{l}{6.309}                            \\
		\multicolumn{2}{l}{CBC}    & \multicolumn{2}{l}{136.800}                          \\
		\multicolumn{2}{l}{CQM}    & \multicolumn{2}{l}{371.975}                         \\		
		\hline	
	\end{tabular}
	\caption{The averaged overall runtime by the CBC, Gurobi (classical) and CQM (quantum) solvers.}
	\label{tab:averaged-overall-runtime}
\end{table}

Investigating the experimental evidence, we can conclude that the classical solvers outperform the CQM solver. Nevertheless, CQM has the potential to be a promising tool for large problems in near future. 

\section{Conclusion and future work}\label{sec:conclusion}

In this paper, we proposed a greedy algorithm for solving the optimization problem of the production of test vehicles using the new hybrid Constrained Quadratic Model Solver (CQM Solver) by D-Wave. We provided a constrained quadratic model formulation for the problem that requires number of qubits linearly proportional to the number of vehicles, car types, features, and tests. We implemented the code for generating the constrained quadratic model and solved the problem instance provided by BMW Quantum Computing Challenge on D-Wave CQM Solver. We benchmarked the results by implementing the integer linear program formulation and running the same algorithm using classical solvers like CBC and Gurobi. The results showed that CQM Solver is comparable to classical solvers in optimizing the number of required vehicles, yet the classical solvers provide the solutions faster. Keeping in mind that the challenges faced and the ongoing efforts in the development of quantum computers, the results obtained from CQM solver for a real-world problem are promising.

The current CQM Solver is limited to 5000 variables, while the real-world problems which are not tractable for classical solvers often require more than that. It is clear that the problem size is an important factor as the currently available quantum solvers are limited in the number of qubits. There are several works that try to formulate models that are more efficient in the number of qubits used \cite{glos2020space, tabi2020quantum, campbell2021qaoa, mohammadbagherpoor2021exploring} and further research can be pursued in this direction. 

A related and more general problem is product configuration and reconfiguration, where the problem's scope is not restricted to vehicles and one may consider any product such as computer parts. Satisfiability based approaches have been considered in \cite{singh2013generation,walter2014remax}. The presented model can be extended for such problems and evoke potential use-cases for D-Wave CQM Solver.

\section*{Acknowledgements}
We would like to thank Jarosław Miszczak and Krzysztof Domino for the discussion on the subject and their valuable comments on the report. This work has been partially supported by Polish National Science Center under the grant agreement 2019/33/B/ST6/02011. AG has been also supported by Polish National Science Center under the grant agreements 2020/37/N/ST6/02220. We would like to thank the organizers of \textit{BMW Group Quantum Computing Challenge} for providing us with the exemplary dataset used in this manuscript

\bibliographystyle{splncs04}
\bibliography{report_bmw}

\end{document}